\documentclass{emulateapj}

\input{colordvi.tex}

\slugcomment{}

\shorttitle{Evolution of the AGN space density}
\shortauthors{Enoki et al.}

\begin{document}

\title{Anti-hierarchical evolution of the Active Galactic Nucleus space density in a hierarchical
universe}

\author{Motohiro Enoki \altaffilmark{1}}
\author{Tomoaki Ishiyama \altaffilmark{2}}
\author{Masakazu  A. R. Kobayashi \altaffilmark{3}}
\author{Masahiro Nagashima \altaffilmark{4, 5}}

\altaffiltext{1}{Faculty of Business Administration, Tokyo Keizai University, Kokubunji, Tokyo, 185-8502, Japan;
enokimt@tku.ac.jp}

\altaffiltext{2}{Center for Computational Sciences,  University of Tsukuba,
Tsukuba, Ibaraki, 305-8577, Japan}

\altaffiltext{3}{Research Center for Space and Cosmic Evolution, Ehime University,
Matsuyama, Ehime, 790-8577, Japan}

\altaffiltext{4}{Faculty of Education, Nagasaki University, Nagasaki, Nagasaki
852-8521, Japan}
\altaffiltext{5}{Faculty of Education, Bunkyo University,
Koshigaya, Saitama, 343-8511, Japan
}

\begin{abstract}
Recent observations show that the space density of luminous active
 galactic nuclei (AGNs) peaks at higher redshifts than that of faint
 AGNs. This downsizing trend in the AGN evolution seems to be contradictory
 to the hierarchical structure formation scenario. In this study, we
 present the AGN space density evolution predicted by a semi-analytic
 model of galaxy and AGN formation based on the hierarchical structure
 formation scenario.
We demonstrate that our model can reproduce the downsizing trend of the
 AGN space density evolution. The reason for the downsizing trend in our model is 
a combination of the cold gas depletion as a consequence of star
formation, the gas cooling suppression in massive halos and the AGN lifetime scaling with the dynamical timescale.
We assume that a major merger
 of galaxies causes a starburst, spheroid formation, and cold gas accretion
 onto a supermassive black hole (SMBH). We also assume that this cold
 gas accretion triggers AGN activity. Since the cold gas is mainly
 depleted by star formation and gas cooling is suppressed in massive
 dark halos, the amount of cold gas accreted onto SMBHs
 decreases with cosmic time. Moreover, AGN lifetime increases with
 cosmic time.
Thus, at low redshifts, major mergers do
 not always lead to luminous AGNs. Because the luminosity of AGNs
 is correlated with the mass of accreted gas onto SMBHs, the space
 density of luminous AGNs decreases more quickly than that of faint AGNs. 
We conclude that the anti-hierarchical
 evolution of the AGN space density is not contradictory to the
 hierarchical structure formation scenario.  
\end{abstract}

\keywords{galaxies: active -- galaxies: evolution -- galaxies: formation --
galaxies: nuclei -- quasars: general}

\section{Introduction}\label{sec:intro}

The space density evolution of active galactic nuclei (AGNs) provides
important clues for understanding of physical processes of AGNs and supermassive black hole (SMBH) evolution.
In the present day universe, massive and spheroidal galaxies are generally
accepted to host SMBHs in the mass
range of $10^6-10^9~M_{\odot}$ in their centers. The masses of SMBHs
are observationally found to correlate with physical properties of the
spheroidal component of the host galaxies 
such as stellar mass and velocity dispersion
\citep[e.g.,][]{Magorrian98, Ferrarese00, Gebhardt00, Marconi03,
Haring04, Gultekin09, Graham12, McConnell13, Kormendy13}. These
relations suggest that the evolution of SMBHs and AGNs is physically
linked to the evolution of galaxies that harbor them. Therefore, in
order to study the AGN space density evolution, it is necessary
to construct a model that includes prescriptions both for galaxy formation
and SMBH/AGN formation. 

The current standard galaxy formation paradigm is 
 the hierarchical structure formation scenario in a cold dark
matter (CDM) universe. In this scenario,
dark halos cluster gravitationally and merge together. 
In each of dark halos, a galaxy is formed at first.
When dark halos merge, the newly formed dark halo contains several galaxies.
These galaxies residing in the same host dark halo can merge together due to
dynamical friction or random collision.
Therefore, more massive galaxies preferentially form at
lower redshifts. Thus, massive SMBHs preferentially form at late
times as the mass of
SMBHs is correlated with the spheroidal stellar component of their
host galaxies.
If the AGN luminosity were directly correlated with the SMBH mass, bright AGNs
would be expected to appear mostly at low redshifts.
However, recent optical observations show that the space densities
of faint AGNs peak at lower redshifts than those of bright AGNs
\citep[e.g.,][]{Croom09, Ikeda11, Ikeda12}. This behavior is called
{\it downsizing} or {\it anti-hierarchical} evolution of AGNs. 
X-ray observations of AGNs also show the
downsizing trend \citep[e.g.,][]{Ueda03, Hasinger05}.
This anti-hierarchical evolution of the AGN density seems to conflict with the hierarchical structure
 formation scenario.

In order to explore the evolution of SMBHs/AGNs, 
many semi-analytic (SA) models of galaxy and AGN formation based on 
the hierarchical structure formation scenario have been proposed 
\citep[e.g.,][]{Kauffmann00, Kauffmann02, Enoki03, Enoki04, Cattaneo05,
 Bower06, Croton06a, Fontanot06, Monaco07, Somerville08}.
Several recent SA models managed to reproduce the downsizing evolution of
AGNs \citep[e.g.,][]{Marulli08, Bonoli09, Fanidakis12, Hirschmann12, Menci13}. 
All of them have changed their models to reproduce the downsizing trend
of AGN evolution. 
For example, \cite{Hirschmann12} modified the SA model of \cite{Somerville08} by considering several
 physical recipes for SMBH growth. \cite{Menci13} used a SA model of galaxy
 and AGN formation in a warm dark matter (WDM) cosmology instead
 of the standard $\Lambda$CDM cosmology. Some recent cosmological
 hydrodynamic simulations of galaxy and AGN evolution can also reproduce
the downsizing trend
of AGN evolution \citep[e.g.,][]{Degraf10, Hirschmann14, Khandai14}. 
Although cosmological hydrodynamic simulations can treat the dynamics of gas components, the
mass resolution is still insufficient to capture the physical origin of
the processes on small scales such as, e.g., star formation, supernova
feedback and gas accretion onto SMBHs.
These simulations nevertheless seem to be able to capture the essence of
how SMBHs grow in reality.

 Recent observations of galaxies also show the
downsizing behavior of galaxy evolution \citep[see the overview of
observations of][and references therein]{Fontanot09}.
Many SA models and cosmological hydrodynamic
 simulations also do not correctly reproduce the
downsizing trends of galaxy evolution.
In particular, SA models and simulations tend to predict
 the larger space density of low-mass galaxies than observed density at
 any redshifts. This is  because low-mass galaxies are
 formed too early in SA models and simulations.
The overestimation of the space density of low-mass galaxies is a major challenge
 to most of current SA models and cosmological hydrodynamic
 simulations \citep[e.g.,][]{Fontanot09, Cirasuol10, Bower12,
 Weinmann12, Guo13}. However, \cite{Henriques13} managed to reproduce the
observed space density evolution of low-mass galaxies due to a combination of
 stronger supernova feedback and delayed re-accretion of gas ejected
 from galaxies by supernova feedback. 

In this paper, we focus on investigating whether or not the observed anti-hierarchical trend of
 the  space density evolution of AGNs is contradictory
 to the hierarchical structure formation scenario.
To this end, we use a SA model of galaxy and SMBH/AGN formation.
Our SA model is the updated {\it Numerical Galaxy
Catalog} \citep[$\nu$GC; ][]{Nagashima05}. $\nu$GC is a SA model based on     
 the Mitaka SA model \citep{Nagashima04} combined  with a  cosmological $N$-body
simulation. In this study, we extend $\nu$GC to incorporate a SMBH/AGN formation model
 of \citet{Enoki03, Enoki04} and use a new $N$-body simulation with
 large box size of \citet{Ishiyama09, Ishiyama12}
 based on the {\it WMAP7} cosmology \citep{Komatsu11}.  
In contrast to other recent SA models, we adopted a simpler
 phenomenological SMBH/AGN formation model, which is a purely major
 merger-driven AGN model, and do not modify our
 existing SA model, $\nu$GC, to reproduce the downsizing trend.

This paper is organized as follows. In Section \ref{sec:model}, we
 briefly describe our SA model, and in Section \ref{sec:evolution}, we present the result of
 the AGN space density evolution. Finally, in Section \ref{sec:conclusion}, we discuss
 the result of our model.

\section{SA model}\label{sec:model}

In SA models, merging histories of dark halos are realized using a
Monte Carlo method or cosmological $N$-body simulations. 
On top of the formation of dark halos, the evolution of baryonic
 components is calculated using simple analytic models for gas cooling,
 star formation, supernova feedback, galaxy merging, and other processes.
In this section, we briefly review our SA model, the $\nu$GC.
Detailed model descriptions are given in \cite{Nagashima05} for galaxies, and
\cite{Enoki03, Enoki04} for SMBH/AGN formation. 

\subsection{Galaxy Formation Model}\label{subsec:galaxy_model}

In our SA model, the merging histories of dark halos are directly taken
from a new large cosmological $N$-body simulation.
It contains $2048^3$ particles within a co-moving
box size of $280~h^{-1}~{\rm Mpc}$.
The gravitational softening length is $4.27~h^{-1}~{\rm kpc}$. 
The particle mass is $1.93 \times 10^8~h^{-1}~M_{\odot}$. 
We used the friends-of-friends algorithm \citep{Davis85} with a linking length parameter of $b = 0.2$ to identify dark halos.
The minimum number of particles identifying a dark halo is 40 and hence
the minimum mass of dark halo is $7.72 \times 10^9~h^{-1}~M_{\odot}$. For the
time integration, we used the GreeM code \citep{Ishiyama09, Ishiyama12}, which is a
massively parallel TreePM code. The cosmological parameters adopted are
based on the concordance $\Lambda$CDM cosmological model \citep[{\it WMAP7}; ][]{Komatsu11}, that is, $\Omega_0=0.2725$, $\lambda_0=0.7275$, $h=0.702$,
 $\Omega_{\rm b} = 0.0455$, $\sigma_8=0.807$, and $n_{\rm s}=0.961$.

When a dark halo is formed, the gas in the halo is shock-heated to the
virial temperature of the halo. We refer to this heated gas as the {\it hot
gas} and provide its radial distribution from the halo center as
isothermal. The hot gas in central dense regions of the halo simultaneously cools due
to efficient radiative cooling. It sinks to the center of the halo and
settles into a rotationally supported disk until the subsequent merger
of the dark halo. 
We refer to this cooled gas as the {\it cold gas}, from which stars can form. The cold gas and stars constitute
{\it galaxies}. In order to avoid the formation of unphysically large
galaxies, the gas cooling process is applied only to dark halos with
circular velocity $V_{\rm circ} \leq V_{\rm cut}$. In this study, we set
$V_{\rm cut} = 260~{\rm km~s}^{-1}$. The halo mass for
cooling suppression is $M_{\rm halo} > 5.77 \times
10^{12} \left( V_{\rm cut} / 260~{\rm km~s}^{-1}\right)^{3}
\left[\Delta_{\rm vir}(z)/200 \right]^{-1/2} M_{\odot}$; 
$\Delta_{\rm vir}(z)$ is the ratio of the dark halo density at redshift
$z$ to the
present critical density. This cooling
suppression is similar to the {\it halo quenching} \citep{Dekel06, Cattaneo08}.

For the star formation rate (SFR) of a galaxy,
we adopt the following form
\begin{equation}
 \psi (t)=\frac{M_{\rm cold}}{\tau_{*}}, \label{eq:SFR}
\end{equation}
where $M_{\rm cold}$ is the mass of the cold gas.
$\tau_{*}$ is the timescale of star formation, which is determined by
\begin{equation}
 \tau_{*} = \tau_{*}^{0} \left[1 + \left(\frac{V_{\rm d}}{V_{\rm hot}}  \right)^{\alpha_{*}} \right], \label{eq:taustar} 
\end{equation}
where $V_{\rm d}$ is the disk rotation velocity.
The two free parameters, $\tau_{*}^{0}$ and $\alpha_{*}$,
are chosen to match the observed mass fraction of cold gas in the neutral
form in the disks of spiral galaxies. 
In this study, we adopted $\tau_{*}^{0} = 1.0~{\rm Gyr} $ and $\alpha_{*} = -5.0$.
$V_{\rm hot}$ is a free parameter related to supernova feedback described below.
When star formation takes place, as a consequence, supernovae
explosions happen and heat up the
surrounding cold gas to the hot gas ({\it supernova feedback}).
The reheating rate of cold gas is given by
 $\dot{M}_{\rm hot} = \beta_{\rm d} \psi (t)$,
 where $\beta_{\rm d}$ is the efficiency of reheating due to supernova
 feedback in disk.
We assume that $\beta_{\rm d}$ depends on $V_{\rm d}$, as follows: 
\begin{equation}
 \beta_{\rm d} = \left(\frac{V_{\rm d}}{V_{\rm hot}}
 \right)^{-\alpha_{\rm hot}}\,. \label{eq:feedback}
\end{equation}
The free parameters of $V_{\rm hot}$ and $\alpha_{\rm hot}$ are determined 
by matching the observed local luminosity function of galaxies. 
In this study, we adopted $V_{\rm hot} = 100~{\rm km~s}^{-1}$ and
$\alpha_{\rm hot} = 4$.

When several dark halos have merged, a newly formed larger dark
halo contains at least two or more galaxies that had originally resided in the individual progenitor halos.  
We identify the central galaxy in the newly formed halo with the central
galaxy contained in the most massive progenitor halo.
Other galaxies are regarded as {\it satellite galaxies}.
These satellites can merge by either dynamical
friction or random collisions.
Satellite galaxies merge with the central
galaxy in the dynamical friction timescale given by \citet{BT}. 
Satellite galaxies sometimes merge with other satellites in 
the timescale of random collisions. 
Under the condition that the satellite galaxies are gravitationally
bound with each other 
and merge during encounters, this timescale is given by \citet{Makino97}. 
Consider the case that two galaxies of masses $m_1$ and $m_2 ~(>m_1)$
merge together. 
If the mass ratio, $f=m_1/m_2$, is larger than a certain
critical value of $f_{\rm bulge}$, we assume that a starburst occurs.
The entire cold gas
content in the progenitor galaxies either turns into bulge stars, is
heated up by supernova feedback, or is accreted
onto the SMBH.
All of the stars pre-existing in merging progenitors and newly formed via
starburst are assumed to populate the bulge of a new galaxy. We call such an
event a {\it major merger}.
On the other hand, if $f<f_{\rm bulge}$, no starburst occurs and a smaller 
galaxy is simply absorbed into the bulge of a larger galaxy. Such an
event is called a {\it minor merger}. 
In this paper, we adopt $f_{\rm bulge} = 0.4$.

Predictions from $\nu$GC are in good agreement with
many observations of, e.g., luminosity functions of local galaxies, cold gas
mass-to-stellar luminosity ratio of spiral galaxies, \ion{H}{1} mass
functions, galaxy sizes, faint galaxy number counts, redshift
distributions for galaxies and the cosmic star formation histories from
$z=0$ to $z=5$ \citep{Nagashima05}. 
However $\nu$GC
does not correctly reproduce the downsizing evolution of the space density of galaxy.
$\nu$GC over-predicted the space density of low-mass galaxies.
As described in Section \ref{sec:intro},
many other SA models and cosmological hydrodynamic simulations also do not correctly reproduce the
downsizing trends of galaxy evolution. Therefore,
this is a common problem in galaxy formation models and simulations based on 
the hierarchical structure formation scenario, suggesting a lack of our understanding on the
interplay between star formation, supernova feedback, gas
recycling, and so on. In this paper, we focus only on the downsizing
evolution of the AGN space density.

\subsection{SMBH/AGN Formation Model}\label{subsec:SMBH_model}

In our model, the SMBH mass growth is assumed to be triggered only by merger events. 
We adopt two processes of SMBH mass growth: (1) SMBH coalescence and
(2) accretion of cold gas during major merger of galaxies. 
We define a SMBH which is powered during a major merger of
galaxies as an AGN. 

\begin{figure}
\epsscale{1.0}
\plottwo{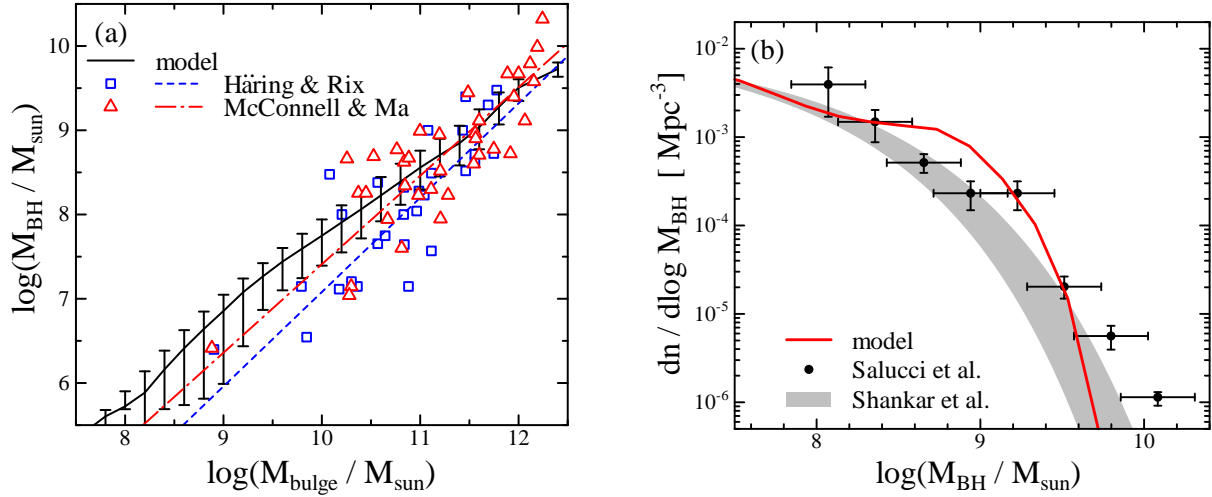}{fig1b.eps}
\caption{
Comparison of our SA model results to observations for SMBHs of the presentday universe.
(a) Bulge -- SMBH mass relation. The solid line indicates the median 
 model results. The ranges represented by vertical solid lines indicate
 the $10 - 90$th percentiles. Observational data are represented by squares 
 \citep{Haring04} and triangles \citep{McConnell13}.
Best-fit power-law relations to the observational
  data are also provided as  dashed \citep{Haring04} and
 dot-dashed \citep{McConnell13} lines.
 (b) Black hole mass function. The solid
 line is the model. Observational data are represented by the symbols 
 \citep{Salucci99} and the shaded area \citep{Shankar04}.
\label{fig1}}
\end{figure}

First, we consider the processes of SMBH mass growth. 
We assume that SMBHs grow through the coalescence of pre-exiting SMBHs when their host galaxies merge.
Some hydrodynamic simulations have shown that a major merger of galaxies can
drive substantial gaseous inflows and fuels a nuclear starburst leading
to the formation of a bulge 
\citep[e.g.,][]{Mihos94, Mihos96, Barnes96, DiMatteo05, Hopkins05, Hopkins06}.
Thus, we also assume that during a major
merger, a fraction of the cold gas, which is  proportional to the total mass of stars newly formed
at the starburst, is accreted onto the SMBH. 
All of the cold gas in the progenitor galaxies is depleted by star
formation, supernova feedback, or accretion on the SMBH.
Under this assumption, the mass of cold gas accreted on the SMBH is given by 
\begin{eqnarray} 
 M_{\rm acc} &=& f_{\rm BH} \Delta M_{*, \rm burst}, \label{eq:bhaccret}
\end{eqnarray} 
where $f_{\rm BH}$ is a constant and $\Delta M_{*, \rm burst} $ is the
total mass of stars formed during the starburst. We set $f_{\rm BH} = 0.0067$ to
 match the observed relation between masses of host bulges and
SMBHs at $z = 0$ found by \citet{Haring04} and \citet{McConnell13}.
Figure \ref{fig1}(a) shows the bulge--SMBH mass relations of
our SA model and observations. 
The analytic description for gas, star and SMBH evolution during a starburst event including  $\Delta M_{*, \rm burst} $  is described in the Appendix.
When a galaxy experiences a major merger for the first time, we assume
that the seed SMBH is formed with mass, $M_{\rm acc}$, which is given
by Equation (\ref{eq:bhaccret}).
As shown in Figure \ref{fig1}(b), the presentday BH mass function predicted by our
SA model, is consistent with the observational results of
\citet{Salucci99} and  \citet{Shankar04} although there
seems to be an excess of $10^{9} M_{\odot}$ black holes. Our galaxy
formation model includes the dynamical friction and the random collision as
galaxy merging mechanisms. The mass function for low-mass black holes is
determined by the random collision between satellite galaxies. The mass
function for high-mass black holes is influenced by the dynamical friction. The
bump at  $\sim 10^{9} M_{\odot}$ of the mass function originates from
the dynamical friction \citep{Enoki04}.

Next, we consider the light curve of AGNs. The accretion of cold gas
during a starburst leads to AGN activity.
Following \citet{Kauffmann00, Kauffmann02}, 
we assume that a fixed fraction of the rest mass energy of the accreted
gas is radiated in the $B$ band, and we adopt the $B$ band
 luminosity of an AGN at time $t$ after a major merger as follows: 
\begin{equation} 
 L_{B}(t) =  \frac{\epsilon_{B} M_{\rm acc} c^2}{t_{\rm life}} \exp(-{\it t}/ {\it t}_{\rm life}) \label{eq:qso-lc},
\end{equation}
where $\epsilon_{B}$ is the radiative efficiency in the $B$-band, 
$t_{\rm life}$ is the AGN lifetime and $c$ is the speed of light. 
In our model, we allow for super-Eddington accretion.
We assume that $t_{\rm life}$ scales with the dynamical timescale, $t_{\rm dyn}$,
as expected if the radius of the accretion disk were to scale with the
radius of the host galaxy \citep{Kauffmann00}.
 The AGN lifetime, $t_{\rm life}$, depends on the redshift of the major merger, since 
$t_{\rm life}  \propto t_{\rm dyn} \propto \rho_{\rm gal}^{-1/2} \propto
 \Delta_{\rm vir}^{-1/2}(z)$; 
$\Delta_{\rm vir}(z)$ is the ratio of the dark halo density at the
redshift of the major merger to the present critical density. 
 In order to determine the parameter
$\epsilon_{B}$ and the present AGN lifetime, $t_{\rm life}(z = 0)$,   
we chose them to match the estimated luminosity function in our model with the
observed $B$-band luminosity function of AGNs at $z=2$. We obtain
$\epsilon_{B} = 0.0055$ and $t_{\rm life}(z = 0)= 5.0 \times 10^{7}~{\rm
yr}$. 

Some recent SA models include prescriptions for the {\it radio-mode
 AGN feedback}. The radio-mode AGN feedback is due to the radio jet
 which is powered by hot gas accretion and injects energy into hot gas. Since the radio-mode AGN feedback is a mechanism
to quench cooling of hot gas and prevent star formation in massive dark
 halos, the radio-mode AGN
feedback is expected to explain part of the downsizing evolution of galaxies
\citep[e.g.,][]{Croton06a, Bower06}. However, the detailed physics of hot gas heating
is still unknown. Moreover, the radio-mode AGN feedback helps to reproduce
 the observed density of massive galaxies but has hardly
 any effects on the space density of low-mass galaxies, which are
 affected by supernova feedback.
Our model does not implement the radio-mode AGN feedback. Instead we take a
simple approach to suppress gas cooling as described in Section \ref{subsec:galaxy_model}.
 The gas cooling process 
is applied only to halos with circular velocity $V_{\rm circ} \leq V_{\rm cut}$.

\section{Evolution of AGN Space Density}\label{sec:evolution}

In Figure \ref{fig2}, we present the redshift evolution of AGN space
density. 
 Solid lines are results of our SA model and symbols
 with error bars linearly connected with dashed lines are results of
 observations \citep{Croom09, Fontanot07, Ikeda11, Ikeda12}.
In order to compare the model result with observations, we convert the absolute $B$-band magnitude ($M_{B}$) into
the absolute AB magnitude at $1450$ \AA ~($M_{1450}$) using the
empirical relation given by \cite{Shen12}. 
Our SA model can qualitatively reproduce the downsizing trend of the AGNs
space density evolution.
While our model overproduces the space density of faint AGN
(i.e. $M_{1450} > -24$) at $ z \lesssim 1$, 
 it is consistent with the results of GOODS
 (\citealp{Fontanot07}) and COSMOS (\citealp{Ikeda11, Ikeda12}) at $z \gtrsim 3$. 

In the following, we will discuss the physical reasons responsible for
the AGN downsizing trend predicted by our SA model.
In our model, SMBHs assemble their mass through cold gas accretion
during a major merger and through the coalescence of the two SMBHs themselves.
At high redshifts, since galaxies generally contain a larger amount of cold
gas, a large amount of the cold gas fuel is
provided for being accreted onto a SMBH.
The cold gas in a galaxy is depleted over time by
star formation. Moreover, the hot gas cooling process is suppressed in massive
dark halos.
Therefore, the amount of the cold gas
 accreted onto SMBH decreases with cosmic time.
Figure \ref{fig3} shows the redshift evolution of the logarithmic mean of the
ratio of cold gas mass ($M_{\rm cold}$) to stellar mass ($M_{\rm star}$) for 
 galaxies with $M_{B} < -18$. This shows that the mass fraction of cold gas mass in a galaxy
 decreases with time and that massive
 galaxies with massive black holes have a smaller cold gas fraction than
 less massive galaxies with less massive black holes. 
 Therefore, at low redshifts, a major merger does not necessarily trigger a luminous AGN because the
luminosity of AGNs is proportional to the cold gas mass of
 accreted onto SMBHs. As a result, the space density of
 luminous AGNs decreases quickly. 
In addition, \cite{Enoki04} showed that at $z \lesssim 1$, a dominant
process for the mass growth of SMBHs is due to black hole coalescence
and not due to gas accretion, since the cold gas decreases with time.

As described in Section \ref{subsec:SMBH_model}, we allow for
super-Eddington accretion. 
Although we do not calculate bolometric luminosity $L_{\rm bol}$, observational studies
have indicated that  $L_{\rm bol}$ is smaller than  $10 L_{B}$. 
Therefore, adopting  $L_{\rm bol} = 10 L_{B}$, we can obtain the upper limit of the number fraction of super-Eddington
AGN ($L_{\rm bol}/ L_{\rm Edd} > 1$) from the $B$-band Eddington ratio
($L_{B}/L_{\rm Edd}$), which can be
calculated in our model.  As shown in Figure \ref{fig4}(a), the 90th percentiles of
$L_{B}/L_{\rm Edd}$ is less than unity at $z < 1$ and larger than unity
at $1 < z <4$, which
results in the fraction of the super-Eddington AGN of less than $10 \%$ at $z < 1$ and
$10 \sim 15\%$ at $1 < z < 4$.  The maximum values of $L_{B}/L_{\rm Edd}$ also indicate that
the maximum Eddington ratio is less than $10$ at $z < 1$ and $10 \sim 80$ at $1 < z < 4$.
The median of the Eddington-ratios of AGNs and 
the fraction of high Eddington ratio AGNs ($\log [L_{B} / L_{\rm Edd}] > -1$) decrease with cosmic time. 
These results are consistent with the observational results of \cite{Shen12}
 and \cite{Nobuta12}.
These trends indicate that the ratio of $\dot{M}_{\rm BH}$ to $M_{\rm BH}$
decreases with time. This suggests that the
 fraction of AGNs having massive black holes in galaxies that have small
 cold gas mass becomes larger at lower redshifts. 

Figure \ref{fig4}(b) shows the redshift evolution of the mean of the logarithm of
 the $B$-band Eddington ratios ($\langle \log [L_{B} / L_{\rm Edd}] \rangle$) in different magnitude intervals. 
We find the following from this figure. (1) The Eddington ratio of
 luminous AGNs is higher than that of faint ones, (2) the Eddington
 ratio of luminous AGNs is almost constant against redshift, and (3) the
 Eddington ratio of faint AGNs decreases with cosmic time. 
In our model, the luminosity of AGNs is
 determined by Equation (\ref{eq:qso-lc}), in which it depends on the
 cold gas mass, not on the SMBH mass.
Thus only AGNs that have large amount of cold gas can be luminous.  
The amount of cold gas generally decreases with cosmic time, while the
 SMBH mass increases with cosmic time.
Therefore, the Eddington ratio of faint AGNs decreases with cosmic time.

In our model, we assume that the AGN lifetime scales with the dynamical
timescale of the host galaxy ($t_{\rm life} \propto t_{\rm dyn}$). 
Since $t_{\rm dyn}$ is shorter at higher redshifts, 
this assumption is one
of the reasons for the AGN downsizing trend.
In order to examine this effect, we plot the redshift evolution of AGN space density
adopting the constant AGN lifetime model for $t_{\rm life} = 5.0 \times 10^{7}~{\rm yr}$
in Figure \ref{fig5} (a). 
The evolution of AGN space density shows neither anti-hierarchical trend nor
hierarchical trend. Thus, the assumption of $t_{\rm life}(z) \propto
t_{\rm dyn}$ is not the unique cause of the downsizing evolution of AGNs.

In many previous studies, the radio-mode AGN
feedback is expected to provide one important process for
predicting the downsizing trend in the space density evolution of
galaxies, as it is supposed to solve the so called {\it over-cooling}
problem and thus to regulate the late growth of massive
galaxies.
 In our model, the gas cooling process is suppressed in dark
halos with circular velocity $V_{\rm circ} > V_{\rm cut}$. This cooling
suppression is similar to the radio-mode AGN feedback.
In order to examine this effect of the gas cooling suppression on the
 evolution of AGN space density, we show the
 evolution of AGN space density of the model for $V_{\rm cut} = 2000~{\rm km~s}^{-1}$ in Figure \ref{fig5}(b).
 In this {\it test} model, the gas cooling process is applied to all
 halos and this corresponds to no radio-mode AGN feedback. Figure \ref{fig5}(b)
 demonstrates that the AGN space density does not show a hierarchical trend. 
Therefore, the cause of the downsizing evolution of AGNs is not only
the radio-mode AGN
feedback, although the radio-mode AGN feedback is an important
ingredient for a galaxy formation model.

In summary, the reason for a downsizing trend of AGN evolution in our
SA model is 
a combination of the cold gas depletion as a consequence of star
formation, the AGN lifetime scaling with the dynamical timescale, and the gas cooling suppression in massive halos.

\section{Conclusion and Discussion}\label{sec:conclusion}

In this study, we have demonstrated that
 our semi-analytic model of galaxy and SMBH/AGN formation based on 
the hierarchical structure formation scenario
predicts a downsizing
trend in the AGN space density evolution in qualitative agreement
with observations. Therefore, the observed anti-hierarchical evolution of the AGN space density is not
 necessarily contradictory to currently favored
 hierarchical structure formation scenarios.
Since the cold gas is mainly depleted by star formation and gas cooling
 is suppressed in massive dark halos, the amount of cold gas accreted
 onto SMBHs decreases with cosmic time. Because the luminosity of AGNs
 is correlated with the mass of accreted gas onto SMBHs, the space
 densities of bright AGNs decrease more quickly than those of faint AGNs. 

Many SA models of galaxy and SMBH/AGN formation have been proposed 
\citep[e.g.,][]{Kauffmann00, Kauffmann02, Enoki03, Enoki04, Cattaneo05, Bower06, Croton06a, Fontanot06, Monaco07, 
Somerville08}. 
Several recent SA models managed to reproduce the downsizing evolution of
AGNs \citep[e.g.,][]{Marulli08, Bonoli09, Fanidakis12, Hirschmann12, Menci13}. 
Unlike these recent SA models, we adopt a simpler
phenomenological SMBH/AGN formation model of \citet{Enoki03, Enoki04},
which is a purely major merger-driven AGN
model.
In our model the anti-hierarchical AGN space density evolution is
a natural result of a combination of different key mechanisms as
discussed in Section \ref{sec:evolution}.
We note that we do not modify our existing model to reproduce the downsizing
trend of the evolution of the AGN space density, in contrast to recent some
 SA models. 
Therefore, reproducing the downsizing trend does not allow for any
conclusion on AGN trigger mechanisms.

In order to determine model parameters related to AGNs, we chose parameters to
match the optical luminosity function in our model with the observed one at $z
 = 2$. The observed luminosity function is not corrected for
dust obscuration effect. X-ray observations of AGNs show that a
significant fraction of AGNs is obscured and the fraction decreases
with X-ray luminosity \citep[e.g.,][]{Ueda03, Hasinger08, Merloni14}.
Therefore, dust obscuration can also strengthen the observed downsizing
trend of AGN evolution.

In our model, we assume that AGN activity is triggered by major
mergers, because major mergers can drive gaseous inflows as
shown in some hydrodynamic simulations as described in Section \ref{subsec:SMBH_model}. 
Recent observations show that the majority of host galaxies of
faint AGNs are  disk-dominated galaxies \citep{Georgakakis09,
Schawinski11},
while bright AGNs appear to be driven by major mergers \citep{Treister12}.  
Therefore, this suggests that it is necessary to include other AGN trigger
mechanisms to explain the evolution of AGNs. However, 
since the cold gas decreases with cosmic time, it is expected that 
the space densities of
bright AGNs decrease more quickly than those of faint AGNs even if we
adopt other AGN trigger mechanisms.

\begin{figure}
\epsscale{0.5}
\plotone{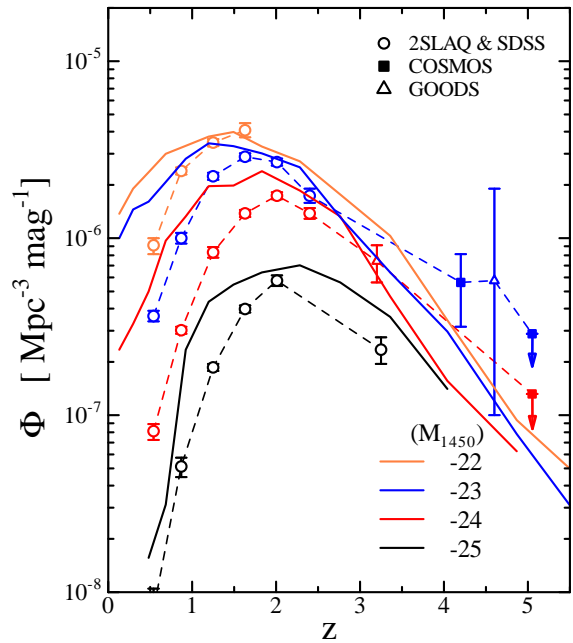}
\caption{
Redshift evolution of the AGN space density. Orange, blue,
 red, and black lines are space density of AGN with $M_{1450} = -22, -23, -24$,
 and $-25$, respectively. Solid lines show model results. Symbols
with error bars linearly connected with dashed lines are observations: 2SLAQ and SDSS (circles;
 \citealp{Croom09}); GOODS (triangles; \citealp{Fontanot07}), and COSMOS
 (filled squares; \citealp{Ikeda11, Ikeda12}). 
\label{fig2}}
\end{figure}

\begin{figure}
\epsscale{0.5}
\plotone{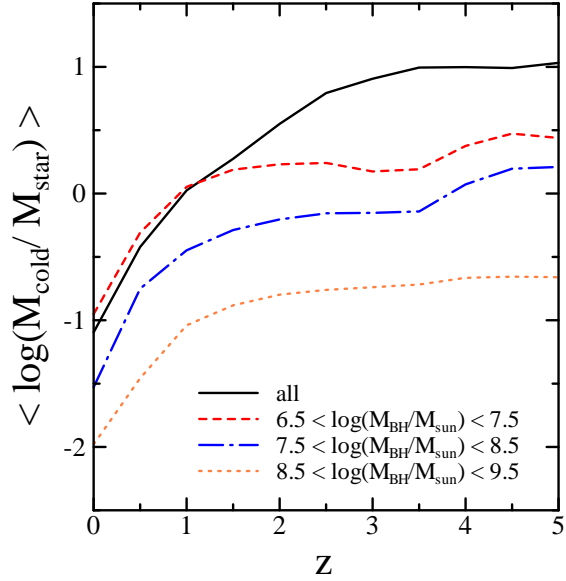}
\caption{
Redshift evolution of the ratios of cold gas mass to stellar mass 
 ($\langle \log [M_{\rm cold} / M_{\rm star}] \rangle$) for 
 galaxies with $M_{B} < -18$. The solid line indicates  the logarithmic
 mean for all  galaxies with $M_{B} < -18$. The dashed, dot-dashed, and
 dotted lines indicate the mean in different SMBH mass intervals 
$6.5  < \log(M_{\rm BH}/M_{\odot}) < 7.5$, $7.5 < \log(M_{\rm
 BH}/M_{\odot}) < 8.5$ and $8.5 < \log(M_{\rm BH}/M_{\odot}) < 9.5$, respectively.  
\label{fig3}}
\end{figure}

\begin{figure}
\epsscale{1.0}
\plottwo{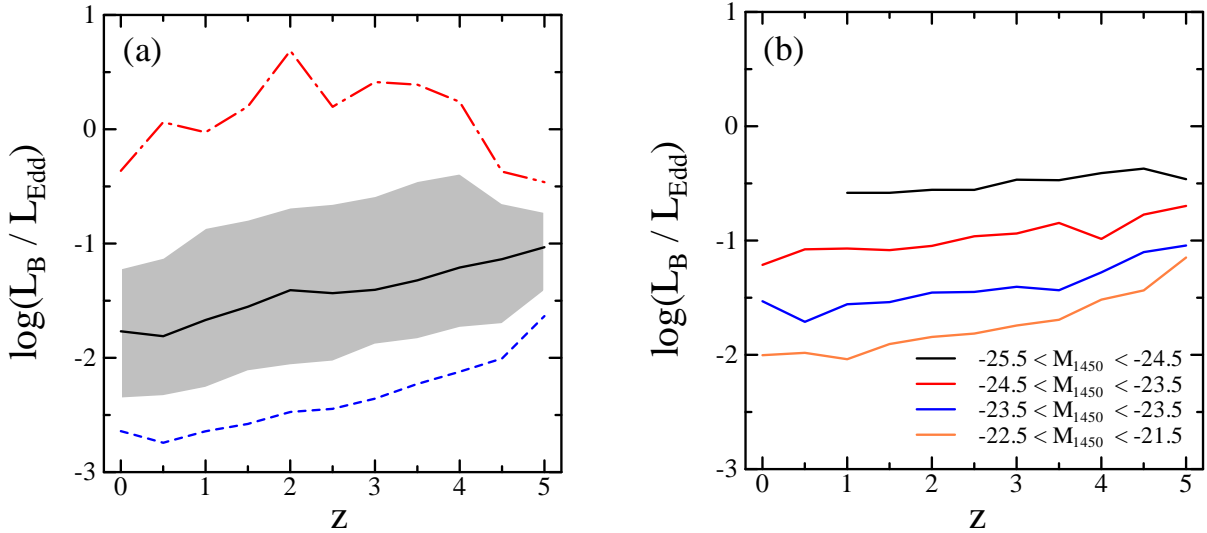}{fig4b.eps}
\caption{
Redshift evolution of the $B$-band Eddington ratios ($\log [L_{B} / L_{\rm Edd}]$) for 
 AGNs with $M_{1450} < -21.5$.
(a) The solid line indicates the median. The shaded area indicates
 $10 - 90$th percentiles. The dot-dashed line shows the maximum value and the
 dashed line shows the minimum value. (b)
 The solid lines indicate the mean in different magnitude intervals 
$ -22.5 < M_{1450} < -21.5$ (orange line),  
$ -23.5 < M_{1450} < -22.5$ (blue line),  
$ -24.5 < M_{1450} < -23.5$ (red line), and 
$ -25.5 < M_{1450} < -24.5$ (black line).
\label{fig4}}
\end{figure}

\begin{figure}
\epsscale{1.0}
\plottwo{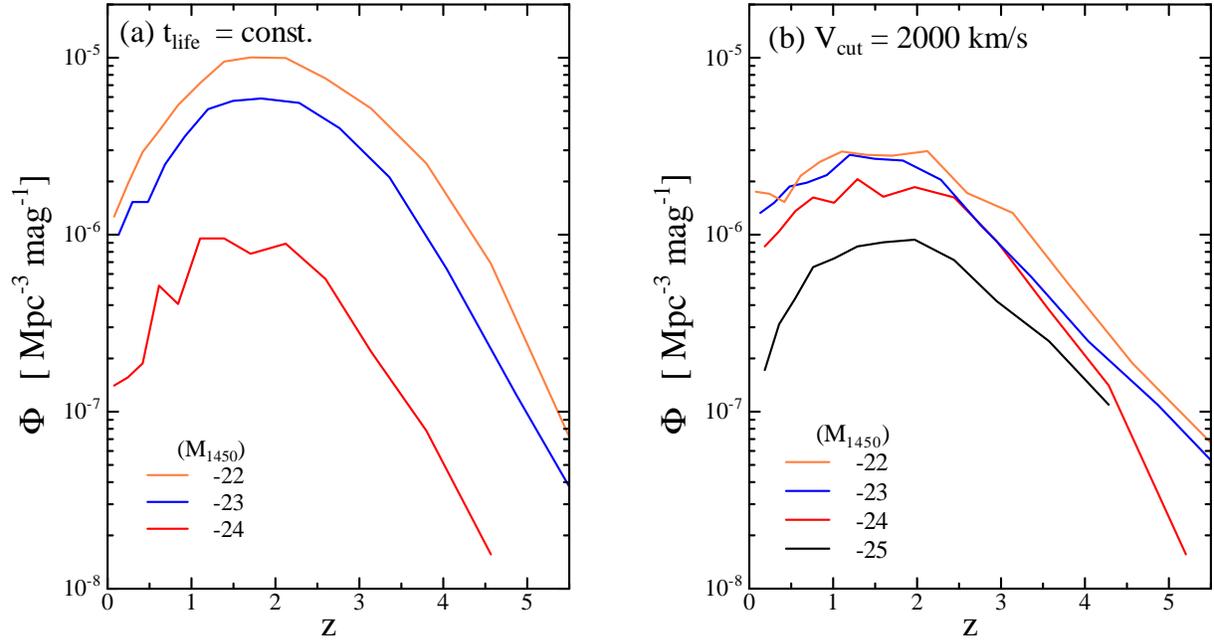}{fig5b.eps}
\caption{
Redshift evolution of the AGN space density of results for (a) the
 constant AGN lifetime model for $t_{\rm life} = 5.0 \times 10^{7}~{\rm
 yr}$ and (b) $V_{\rm cut} = 2000~{\rm km~s}^{-1}$. 
Orange, blue, red, and black lines are space density of AGNs with $M_{1450} = -22, -23, -24$,
 and $-25$, respectively. 
\label{fig5}}
\end{figure}

At low redshifts ($ z \lesssim 1$), the faint AGN space density in
our model is larger than the observed faint AGN density (see Figure \ref{fig2}). This suggests 
that the cold gas mass accreted on a SMBH in our model is too large at
$z \lesssim 1$. In our model, we assume that
during a major merger, all cold gas supplied
from a host galaxy accretes onto a SMBH. However, 
 all the gas that is driven into the central region of the galaxy is not
 accreted onto the SMBH.
Since the angular momentum of cold gas cannot be thoroughly removed,
the cold gas forms a circumnuclear disk in the central $\sim 100~{\rm pc}$ around the SMBH.
\citet{Kawakatu08} proposed an evolutionary model of a SMBH and a
circumnuclear disk. They 
found that not all the cold gas supplied from the host galaxy accretes onto the
SMBH. This is because part of the gas is used to form stars in the circumnuclear disk.
As a result, the final SMBH mass ($M_{\rm SMBH, final}$) is not
proportional to the total cold gas mass supplied from host galaxy ($M_{\rm sup}$)
during the hierarchical formation of galaxies; $M_{\rm SMBH, final} /
M_{\rm sup}$ decreases with $M_{\rm sup}$ increasing. We plan to update our
SA model to include this model. 

At high redshifts ($z \gtrsim 3$), 
there is a discrepancy between observational results of faint AGNs
space densities. 
The results of GOODS (\citealp{Fontanot07}) and COSMOS
(\citealp{Ikeda11, Ikeda12, Masters12}) 
showed that the faint AGN space density decreases with redshift (see
Figure \ref{fig2}). Our model result is consistent with these
observational results.
The results of \citet{Glikman10, Glikman11}, however, showed constant or higher space densities
of AGNs with $M_{1450} \gtrsim -24$. 
Since not all of the observations take dust obscuration correction
into account, the results do not depend on the recipe of
dust correction.  Both of \cite{Ikeda11} and \cite{Masters12} provide a
comment on the discrepancy that the larger space density obtained by
\cite{Glikman10, Glikman11} can be attributed to a large fraction of contaminants such
as Lyman-Alpha Emitters and Lyman-Break Galaxies.
Further observations of faint AGNs
in a wider survey area are crucial to obtain AGN space densities. 
A forthcoming wide-field survey using the {\it Subaru} Telescope with Hyper
Suprime-Cam (\citealp{Miyazaki06,
Miyazaki12})\footnote{http://www.naoj.org/Projects/HSC/index.html} will provide useful constraints on the SMBH and AGN evolution model.

\acknowledgments

We appreciate the detailed reading and useful suggestions of anonymous
referee that have improved our paper. 
We thank H. Ikeda for providing us with the compiled
observational data of the quasar space density.
We also thank N. Kawakatu, K. Wada, T. Nagao, T. R. Saitoh, and S. Ichikawa for
useful comments and discussion.
Numerical computations were partially carried out on Cray XT4 at
Center for Computational Astrophysics, CfCA, of National Astronomical
Observatory of Japan. T.I. is financially supported by MEXT HPCI
STRATEGIC PROGRAM and MEXT/JSPS KAKENHI grant No. 24740115.
M.N. is supported by the Grant-in-Aid (Nos. 22740123 and 25287049) from the Ministry of
Education, Culture, Sports, Science, and Technology (MEXT) of Japan.


\appendix

\section{Gas, Star and SMBH Evolution in a Starburst Galaxy}\label{sec:ap1}
In this appendix, we summarize our model of gas, star
and SMBH evolution during a starburst event. We use a simple
instantaneous recycling approximation model of star formation,
supernovae feedback, and chemical enrichment.
The following 
differential equations describe the evolution of the mass of cold
gas $M_{\rm cold}$, hot gas $M_{\rm hot}$, long lived stars $M_{\rm
star}$ and SMBH $M_{\rm BH}$ at each time step:
\begin{eqnarray}
\dot{M}_{\rm cold} &=& -\alpha \psi(t) -\beta \psi(t) -f_{\rm BH} \psi(t), 
 \label{eq:coldeq} \\
\dot{M}_{\rm hot} &=&\beta \psi(t), \label{eq:hoteq} \\
\dot{M}_{\rm star} &=& \alpha \psi(t), \label{eq:stareq} \\
\dot{M}_{\rm BH} &=& f_{\rm BH} \psi(t), \label{eq:BHeq} 
\end{eqnarray}
where $\psi(t)={M_{\rm cold}}/{\tau_{*}}$ is star formation rate,
 $\alpha$ is a locked-up mass fraction, and $\beta$ is the
efficiency of reheating for starburst. 
 In this paper, we set $\alpha = 0.75$ 
in order to be consistent with a stellar evolution model. 
 $\beta$ is the starburst supernovae feedback strength, which is given by
 $\beta = (V_{\rm b}/V_{\rm hot})^{-\alpha_{\rm hot}}$, 
where $V_{\rm b}$ is the velocity dispersion of the newly formed bulge. 
The solutions of these equations are as follows.
\begin{eqnarray}
M_{\rm cold} &=& M_{\rm cold}^{0} \exp \left[-\left(\alpha+\beta+f_{\rm BH}
 \right)\frac{t}{\tau_{*}} \right], \label{eq:coldsol} \\ 
M_{\rm hot} &=& M_{\rm hot}^{0} + \beta \Delta M_{*}, \label{eq:hotsol} \\
M_{\rm star} &=& M_{\rm star}^{0} + \alpha\Delta M_{*},
 \label{eq:starsol} \\
M_{\rm BH} &=& M_{\rm BH}^{0} + f_{\rm BH} \Delta M_{*}, \label{eq:BHsol} 
\end{eqnarray}
where $t$ is the time since the starburst, $M_{\rm cold}^{0}, M_{\rm hot}^{0}$, $M_{\rm star}^{0}$ and
$M_{\rm BH}^{0}$ are the masses of cold gas, hot gas, stars and SMBH in progenitor galaxies at the initial state $t = 0$, 
and $\Delta M_{*} = (M_{\rm cold}^{0} - M_{\rm cold})/(\alpha+\beta+f_{\rm BH})$ is the mass of total formed stars.  

When a starburst occurs, stars are formed on a very short
timescale. Thus, the starburst corresponds to $t/\tau_{*} \to \infty$ in the above
solutions. Therefore, the changes of masses are given by
\begin{eqnarray} 
M_{\rm cold} &=& 0, \label{eq:coldburst} \\ 
M_{\rm hot} &=& M_{\rm hot}^{0} + \frac{\beta M_{\rm cold}^{0}}{\alpha+\beta+f_{\rm BH}}, \label{eq:hotburst} \\
M_{\rm star} &=& M_{\rm star}^{0} + 
 \frac{\alpha M_{\rm cold}^{0}}{\alpha+\beta+f_{\rm BH}}, \label{eq:starburst} 
\end{eqnarray}
 and the total star mass formed at starburst becomes 
\begin{equation}
\Delta M_{*, \rm burst} = \frac{M_{\rm cold}^{0}}{\alpha+\beta+f_{\rm BH}}. \label{eq:totstar} 
\end{equation}
From Equation (\ref{eq:totstar}), we can obtain the mass of cold
gas accreted onto a black hole [Equation (\ref{eq:bhaccret})].


\clearpage


\end{document}